\documentclass[preprint,showpacs,aps]{revtex4}

\usepackage{graphicx}
\usepackage{dcolumn}
\usepackage{bm}

\begin{document}


\title{Photoassociation of a quantum degenerate gas}

\author{Ionut D. Prodan}
\author{Marin Pichler}
\author{Mark Junker}
\author{Randall G. Hulet}
\affiliation{Department of Physics and Astronomy and Rice Quantum
Institute, Rice University, Houston, Texas 77251}
\author{John L. Bohn}
\affiliation{JILA, National Institute of Standards and Technology
and University of Colorado, Boulder, Colorado 80309-0440}

\date{July 13, 2003}

\begin{abstract}
We have measured the intensity dependent rate and frequency shift
of a photoassociation transition in a quantum degenerate gas of
$^7$Li.  The rate increases linearly with photoassociation laser
intensity for low intensities, whereas saturation is observed at
higher intensities.  The measured rates and shifts agree
reasonably well with theory within the estimated systematic
uncertainties. Several theoretically predicted saturation
mechanisms are discussed, but a theory in which saturation arises
because of quantum mechanical unitarity agrees well with the data.
\end{abstract}


\maketitle

Photoassociation (PA) of ultracold atoms has been a remarkably
useful tool for determining scattering lengths characterizing
ultracold atom collisions, for producing ultracold molecules, and
for providing extremely precise measurements of atomic radiative
lifetimes (see Refs. \cite{Weiner99, Stwalley99} for reviews).
This utility is largely a consequence of the spectroscopic
precision afforded by the small thermal broadening in a laser or
evaporatively cooled gas. Quantum degenerate gases are especially
interesting because the coherence of the atomic field may enable
the formation of a molecular Bose-Einstein condensate (BEC) from
an atomic one by coherent Raman transitions \cite{Julienne98,
Heinzen00}, stimulated Raman adiabatic passage (STIRAP)
\cite{Mackie00}, or by other coherent adiabatic population
transfer schemes \cite{Javanainen99, Mies00}.

There have been extensive theoretical studies on the rate of PA
\cite{Napolitano94, Pillet97, Cote98, Bohn99, Kostrun00}.  The
rate is predicted to increase linearly with intensity at low
intensities, while various mechanisms have been proposed that
cause saturation of the rate at higher intensities.  Among these
mechanisms are the quantum mechanical unitarity limit on the rate
of atomic collisions \cite{Bohn99}, a break-down of the 2-mode
approximation \cite{Timmermans99, Heinzen00} for PA of Bose
condensates caused by coupling to non-condensed atomic modes
\cite{Kostrun00, Goral01, Holland01, Javanainen02}, and the
depletion of the atomic pair correlation function
\cite{Holland01}. Photoassociation resonances are also predicted
to exhibit a spectral shift proportional to the light intensity
caused by coupling to the continuum of free-atom states
\cite{Fedichev96, Bohn97, Javanainen98, Bohn99}.   In contrast to
theory, there are relatively few experimental measurements, and
only two that could be considered ``precise" (which we define as
measurements with uncertainties of less than a factor of 2). We
previously measured the spectral light shift using quantum
degenerate $^{7}$Li and obtained good agreement with theory
\cite{Gerton01}. Both the spectral shift and the PA rate constant
were recently measured in a Na condensate, and good agreement with
two-body theory was found \cite{McKenzie02}.  Saturation was not
observed in this experiment. Saturation was observed in two other
lower precision experiments \cite{Drag00, Schloder02}, but these
experiments were performed in a magneto-optical trap, where the
temperatures were greater than 100 $\mu$K and the corresponding
unitarity-limited rates were quite small, of the order of 1
s$^{-1}$ or less.

We report precise measurements of both the rate of PA and the
spectral shift in a quantum degenerate gas of $^7$Li atoms as a
function of laser intensity.  The rate is observed to saturate at
the highest intensities, where the corresponding rate constants
are nearly two orders of magnitude larger than any previously
measured.  The large PA rate is a consequence of quantum
degeneracy, where the densities are relatively high and the
unitarity limited rates are large, as well as favorable free-bound
state overlap in lithium.

The apparatus will be discussed only briefly as it has been
described in detail previously \cite{Sackett97b, Gerton01}. Atoms
are confined by a magnetic trap with a bias field of $\sim$1000 G.
Atoms in the F = 2, m$_{F} = 2$ hyperfine sublevel of $^{7}$Li are
cooled by evaporation to quantum degeneracy. Attractive
interactions between atoms restrict the number of atoms in the
Bose-Einstein condensate, $N_0$, to a relatively small number
before the condensate collapses \cite{ Bradley97a}. For the
experimental conditions here, $N_0 \lesssim 1250$, a small
fraction of the total number of atoms. This fact, however,
facilitates the achievement of repeatable temperatures, total atom
numbers, and densities because quenching the gas below the BEC
transition temperature $T_c$ causes it to attain thermal
equilibrium with a temperature $T \simeq T_c$ \cite{Sackett99}.
Therefore, following evaporation we allow the gas to equilibrate
for approximately 12 s, at which point the total number of atoms
$N \simeq 8\times10^{5}$, the gas is very near the BEC transition
at $T \simeq 600$ nK, and the peak density is
$\sim$4$\times10^{12}$ cm$^3$.

Following equilibration, the atoms are exposed to a pulse of
photoassociation light.  Up to 85 mW of PA light, coupled out of a
single-mode fiber is focused on to the atoms. The laser frequency
is tuned to near resonance with the $v'=83$ vibrational level of
the $1^3\Sigma_g^+$ state, which has a binding energy of 60 GHz
relative to the 2P$_{1/2}$ asymptote \cite{Abraham95b}. The
relative frequency of the PA laser is monitored by comparing it
with a reference laser locked to an atomic transition using a
scanning Fabry-Perot etalon.  The $v'=83$ level was chosen for its
large ratio of the photoassociation rate to the rate of
off-resonant scattering from the atomic resonance.

Photoassociation can be detected because it causes a reduction in
atom number when excited molecules spontaneously decay into pairs
of hot atoms that escape the trap \cite{ Weiner99, Stwalley99}.
 The number of atoms remaining following the PA pulse is determined
by polarization phase-contrast imaging \cite{ Sackett97b}, and
compared with a ``background'' measurement of the number of atoms
without the PA pulse.  Since imaging destroys the sample, the trap
must be reloaded and the atoms evaporatively cooled for each
image.  Both $N$ and $T$ are extracted by fitting the images to
the distribution function for a Bose gas in an external potential.
The photoassociation rate is determined for a given laser
intensity $I$ by adjusting the laser pulse length $\tau$ to cause
a 20\% - 30\% loss of the initial number of atoms. For the data
presented here, $\tau$ is between 12 $\mu$s at the highest
intensity and 4 ms at the lowest. Other phenomena besides PA, such
as spontaneous scattering from the atomic resonance and dipole
forces, can cause light-induced trap losses. Dipole forces,
arising from gradients of the laser field, are minimized by using
a large laser beam waist (1/$e^{2}$ radius of between 220 and 320
$\mu$m). The background loss rate is measured by tuning the PA
laser several GHz from the molecular resonance, and the data is
adjusted accordingly, by up to 2\% at the highest intensities.
Fig. 1 shows a typical photoassociation resonance curve for
relatively low intensity.  The curve is well-described by a
Lorentzian with a width only slightly larger than the expected
value of twice the atomic natural linewidth. Although we did not
perform a systematic study of lineshapes, the resonances broaden
as expected as $I$ is increased. The resonance position shifts
with laser intensity, as shown in Fig. 2. A linear fit to the data
gives a slope of $(-1.7 \pm 0.2)$ MHz/(W-cm$^{-2}$).

The measurement of the rate is expressed in terms of the
on-resonance rate coefficient $K_{p}$, which is defined in terms
of the time derivative of the density
$\dot{n}(\mathbf{r},t)=-K_{p}(I)n^{2}(\mathbf{r},t)$.  We neglect
the effect of elastic collisions on the density distribution since
$\tau$ is much less than the characteristic time for elastic
scattering of $\sim$300 ms. The solution for the evolution of the
density,
$n(\mathbf{r},t)=n(\mathbf{r},0)/\{1+K_{p}(I)n(\mathbf{r},0)t\}$,
shows that the distribution becomes flatter during the PA pulse
since the rate is largest at the highest densities
\cite{McKenzie02}. The initial density $n(\mathbf{r},0)$ prior to
the PA pulse is described by an equilibrium Bose-Einstein density
distribution. The fractional trap loss, for a given value of $K_p$
and $\tau$, is calculated by taking the ratio between the spatial
integral of $n(\mathbf{r},\tau)$ and the initial number of atoms.
The best value for $K_{p}$ is found numerically by matching the
computed fractional loss with the measured one.  Fig. 3 shows the
dependence of the on-resonance rate coefficient $K_p$ on the laser
intensity $I$.  In the low-intensity limit $K_p$ is found to
depend linearly on the laser intensity, with a slope of $(7.9 \pm
1.6) \times 10^{-10}$ (cm$^3$-s$^{-1}$)/(W-cm$^{-2}$). The rate
constant saturates at high intensities, with a maximum value of
$(2.2 \pm 0.2) \times 10^{-8}$ cm$^{3}$-s$^{-1}$.  The stated
uncertainties are dominated by systematic effects.

Several factors contribute to the uncertainties of the measured
quantities.  There is no direct way to know $N$ before exposure to
the PA pulse and therefore the ability to measure PA-induced loss
depends on the reproducibility of $N$ and $T$. We find that the
statistical variations in $N$ and $T$ for the background
measurements have standard deviations of only 3\% and 2\%,
respectively.  Furthermore, the phase-space density, proportional
to $NT^{-3}$, is found to be within 5\% of the critical value for
all background measurements. Background measurements are typically
made every fourth shot to monitor and make adjustments for
long-term drifts.  Of more concern are systematic uncertainties.
$N$ and $T$ are subject to systematic uncertainties of 5\% and
3\%, respectively, due to uncertainties in the image magnification
and probe laser polarization \cite{Sackett97b}. Finally, the
uncertainty in intensity is dominated by a systematic uncertainty
in power measurement, which we estimate to be $\sim$10\%.  These
effects combine to give a 20\% uncertainty to the slope of $K_p$,
and a 10\% uncertainty to both the rate constant maximum and the
slope of the frequency shift.

The data are compared with the theory of Ref. \cite{Bohn99}. The
transition dipole, $d_m$, for the free-bound transition to this
high-lying vibrational level is simply related to the atomic
dipole $d_a$. Since the binding energy of $v'=83$ is large
compared to the fine-structure splitting of the 2P atomic state
and any hyperfine interactions, the electronic and nuclear spins
of the atoms decouple from the PA light field. Therefore, for this
$\Sigma \rightarrow \Sigma$ transition, $d_m$ lies along the
internuclear axis. Averaging over the unknown orientation of the
molecular axis relative to the polarization axis of the linearly
polarized laser field gives a factor of $\sqrt{1/3}$, so that $d_m
= \sqrt{2/3}d_a$ \cite{footnote}. The free-bound Franck-Condon
factors are numerically evaluated using molecular potentials
determined previously \cite{ Abraham95a, McAlexander95}.  Finally,
because of quantum degeneracy, to calculate $K_p$ we average the
scattering matrix element over a Bose-Einstein energy
distribution.

The light-induced spectral shift is calculated to be $-$2.1
MHz/(W-cm$^{-2}$), in reasonably good, but not perfect agreement
with the measured value.  The solid line in Fig. 3 shows the
theoretical results for $K_p$.  The dashed line is a low-intensity
extrapolation of the theory which has a slope of $5.8 \times
10^{-10}$ (cm$^3$-s$^{-1}$)/(W-cm$^{-2}$), which again is in
reasonable agreement.  Finally, theory predicts saturation to
$K_{p}$ = $2.0 \times 10^{-8}$cm$^3$-s$^{-1}$, which agrees very
well with the measured value.

The theory of Refs. \cite{Napolitano94} and \cite{Pillet97} show
that $K_p$ goes to a constant value, independent of $T$, in the
low-$T$ and low-$I$ limits.  However, the free-bound matrix
element depends on $T$ in this theory.  We find it convenient to
define the free-bound matrix element as $T$-independent from the
start, and use this to obtain a completely $T$-independent
analytic approximation for $K_p$. Therefore, we scale the two-atom
unbound wave function $\phi_k (r)$ such that outside of the
interior of the ground-state molecular potential $\phi_k
(r)=\frac{\sin(kr+\delta)}{kr}$, where $\delta$ is the s-wave
scattering phase shift.  The overlap between the unity-normalized
molecular wave function $\psi_v (r)$ and $\phi_k (r)$ is $M_{kv} =
4\pi\int_0 ^{\infty}r^2 \phi_k (r)\psi_v ^* (r)dr$.
 This matrix element has units of $(length)^{3/2}$ and reaches a
constant value as $T\rightarrow0$.  In our experiment, for
example, at $T \simeq 600$ nK and for $v' = 83$, the wave number
$k \simeq 2 \times 10^{-4} a_o^{-1}$, the classical outer turning
point $R_{c} = 103$ $a_o$ and the \textit{s}-wave scattering
length $a = -27.6$ $a_o$ \cite{Abraham95a}, where $a_o$ is the
Bohr radius. The value of $|M_{kv}|^{2}$ for these conditions
differs from the $T\rightarrow0$ value by only 0.02\%, showing
that the system is well within the quantum threshold regime.  At
low intensities and temperatures, we find
$K_{p}=\gamma\frac{I}{I_{sat}}|M_{kv}|^{2}$, where $I_{sat}$ and
$\gamma$ are the molecular saturation intensity and spontaneous
decay rate, respectively.  For this transition $I_{sat} = 6I_a$,
where $I_a = 5.1$ mW/cm$^2$ is the atomic saturation intensity.

Saturation comes about in the theory of Ref. \cite{Bohn99} because
of the quantum mechanical unitarity limit on the rate of two-body
collisions. Accordingly, the upper bound for the rate constant is
$K_p ^{(u)}=\sigma \langle v \rangle$, where
$\sigma=\frac{2}{\pi}\lambda^{2}$ is the maximum cross section
between identical, non-condensed bosons, $\langle v \rangle$ is
the thermally averaged velocity, and $\lambda$ is the thermal de
Broglie wavelength. For $T = 600$ nK, $K_{p}^{(u)}= 2.5 \times
10^{-8}$ cm$^{3}$ s$^{-1}$, in good agreement with the full
theoretical calculation and experiment. The saturation observed in
previous experiments \cite{Drag00, Schloder02} is consistent with
the unitarity limit, although at rates many orders of magnitude
lower than observed here. It is interesting to speculate on how
our results relate to the other proposed saturation mechanisms.
The reverse process of dissociation can result in the formation of
pairs of ``hot" atoms that are not returned to the original
translational state of the cold atomic gas \cite{Kostrun00,
Goral01, Holland01, Javanainen02}. The energy width of the
dissociated pairs is proportional to the PA rate, and results in a
rate limit of $\sim$$\hbar n^{2/3}/m$ \cite{Javanainen02}.
Interestingly, at $T_c$, where $\lambda \propto n^{-1/3}$, the
unitarity limit gives the same maximum rate $nK_p ^{(u)}$ $\simeq$
$\hbar n^{2/3}/m$. Since association must involve atom pairs in
close proximity, depletion of the pair correlation function at
short range may also limit the rate.  From a classical,
particle-like perspective, the maximum association rate is reached
when atoms cannot move quickly enough to replenish the necessary
close-range pairs, leading to a maximum rate of $\sim$$v/d$, where
$d \simeq n^{-1/3}$ is the mean separation between atoms and $v$
is their mean velocity.  But since $v \simeq \hbar/m\lambda$ and
$n \simeq N_0/\lambda^3$, the resulting limit is
$\sim$$N_0^{-1/3}\hbar n^{2/3}/m$, which is again the same as the
unitarity limit when $T \simeq T_c$.  The limit imposed by this
classical argument, however, has already been violated in the BEC
experiment reported in Ref. \cite{McKenzie02}, in which a PA rate
greatly in excess of $v/d$ was observed. Although no saturation
was observed in that experiment, the rate was well below the
unitarity limit, and $\hbar n^{2/3}/m$.  Our experiment cannot
clearly distinguish between the mechanisms because their predicted
limits are all of the same order at $T_c$. However, the overall
good quantitative agreement over a large range of intensity with a
theory whose rate is limited by unitarity in the two-body
scattering amplitude, strongly supports this explanation for our
data.  Finally, we point out that mechanisms limiting the PA rate
will also limit the rate of molecule formation by magnetic
Feshbach resonances.

In summary, we report the results of precise measurements of the
rate and spectral shift of photoassociation resonances of a
quantum degenerate gas.  The overall agreement between theory,
with no free parameters, and our measurement is good. The
agreement attests to the general validity of the theory and also
points to quantum mechanical unitarity as the limitation of the
rate for a gas at $T_c$.  The distinction between the
unitarity-limited model and the hot dissociation model could
nevertheless be unambiguously tested with a lithium gas cooled
significantly below $T_c$, an experiment we hope to perform in the
future.

We thank Robin C\^{o}t\'{e}, Juha Javanainen and Paul Julienne for
helpful discussions.  This work was partially funded by grants
from the National Science Foundation, the Office of Naval
Research, the NASA, and the Welch Foundation.

\bibliographystyle{apsrev2}

\begin{thebibliography}{30}
\expandafter\ifx\csname
natexlab\endcsname\relax\def\natexlab#1{#1}\fi
\expandafter\ifx\csname bibnamefont\endcsname\relax
  \def\bibnamefont#1{#1}\fi
\expandafter\ifx\csname bibfnamefont\endcsname\relax
  \def\bibfnamefont#1{#1}\fi
\expandafter\ifx\csname citenamefont\endcsname\relax
  \def\citenamefont#1{#1}\fi
\expandafter\ifx\csname url\endcsname\relax
  \def\url#1{\texttt{#1}}\fi
\expandafter\ifx\csname
urlprefix\endcsname\relax\def\urlprefix{URL }\fi
\providecommand{\bibinfo}[2]{#2}
\providecommand{\eprint}[2][]{\url{#2}}

\bibitem[{\citenamefont{Weiner et~al.}(1999)\citenamefont{Weiner, Bagnato,
  Zilio, and Julienne}}]{Weiner99}
\bibinfo{author}{\bibfnamefont{J.}~\bibnamefont{Weiner}},
  \bibinfo{author}{\bibfnamefont{V.}~\bibnamefont{Bagnato}},
  \bibinfo{author}{\bibfnamefont{S.}~\bibnamefont{Zilio}}, \bibnamefont{and}
  \bibinfo{author}{\bibfnamefont{P.~S.} \bibnamefont{Julienne}},
  \bibinfo{journal}{Rev. Mod. Phys.} \textbf{\bibinfo{volume}{71}},
  \bibinfo{pages}{1} (\bibinfo{year}{1999}).

\bibitem[{\citenamefont{Stwalley and Wang}(1999)}]{Stwalley99}
\bibinfo{author}{\bibfnamefont{W.~C.} \bibnamefont{Stwalley}} \bibnamefont{and}
  \bibinfo{author}{\bibfnamefont{H.}~\bibnamefont{Wang}}, \bibinfo{journal}{J.
  Mol. Spectrosc.} \textbf{\bibinfo{volume}{195}}, \bibinfo{pages}{194}
  (\bibinfo{year}{1999}).

\bibitem[{\citenamefont{Julienne et~al.}(1998)\citenamefont{Julienne, Burnett,
  Band, and Stwalley}}]{Julienne98}
\bibinfo{author}{\bibfnamefont{P.~S.} \bibnamefont{Julienne}},
  \bibinfo{author}{\bibfnamefont{K.}~\bibnamefont{Burnett}},
  \bibinfo{author}{\bibfnamefont{Y.~B.} \bibnamefont{Band}}, \bibnamefont{and}
  \bibinfo{author}{\bibfnamefont{W.~C.} \bibnamefont{Stwalley}},
  \bibinfo{journal}{Phys. Rev A} \textbf{\bibinfo{volume}{58}},
  \bibinfo{pages}{R797} (\bibinfo{year}{1998}).

\bibitem[{\citenamefont{Heinzen et~al.}(2000)\citenamefont{Heinzen, Wynar,
  Drummond, and Kheruntsyan}}]{Heinzen00}
\bibinfo{author}{\bibfnamefont{D.~J.} \bibnamefont{Heinzen}},
  \bibinfo{author}{\bibfnamefont{R.}~\bibnamefont{Wynar}},
  \bibinfo{author}{\bibfnamefont{P.~D.} \bibnamefont{Drummond}},
  \bibnamefont{and} \bibinfo{author}{\bibfnamefont{K.~V.}
  \bibnamefont{Kheruntsyan}}, \bibinfo{journal}{Phys. Rev. Lett.}
  \textbf{\bibinfo{volume}{84}}, \bibinfo{pages}{5029} (\bibinfo{year}{2000}).

\bibitem[{\citenamefont{Mackie et~al.}(2000)\citenamefont{Mackie, Kowalski, and
  Javanainen}}]{Mackie00}
\bibinfo{author}{\bibfnamefont{M.}~\bibnamefont{Mackie}},
  \bibinfo{author}{\bibfnamefont{R.}~\bibnamefont{Kowalski}}, \bibnamefont{and}
  \bibinfo{author}{\bibfnamefont{J.}~\bibnamefont{Javanainen}},
  \bibinfo{journal}{Phys. Rev. Lett.} \textbf{\bibinfo{volume}{84}},
  \bibinfo{pages}{3803} (\bibinfo{year}{2000}).

\bibitem[{\citenamefont{Javanainen and Mackie}(1999)}]{Javanainen99}
\bibinfo{author}{\bibfnamefont{J.}~\bibnamefont{Javanainen}} \bibnamefont{and}
  \bibinfo{author}{\bibfnamefont{M.}~\bibnamefont{Mackie}},
  \bibinfo{journal}{Phys. Rev. A} \textbf{\bibinfo{volume}{59}},
  \bibinfo{pages}{R3186} (\bibinfo{year}{1999}).

\bibitem[{\citenamefont{Mies et~al.}(2000)\citenamefont{Mies, Tiesinga, and
  Julienne}}]{Mies00}
\bibinfo{author}{\bibfnamefont{F.~H.} \bibnamefont{Mies}},
  \bibinfo{author}{\bibfnamefont{E.}~\bibnamefont{Tiesinga}}, \bibnamefont{and}
  \bibinfo{author}{\bibfnamefont{P.~S.} \bibnamefont{Julienne}},
  \bibinfo{journal}{Phys. Rev. A} \textbf{\bibinfo{volume}{61}},
  \bibinfo{pages}{022721} (\bibinfo{year}{2000}).

\bibitem[{\citenamefont{Napolitano et~al.}(1994)\citenamefont{Napolitano,
  Weiner, Williams, and Julienne}}]{Napolitano94}
\bibinfo{author}{\bibfnamefont{R.}~\bibnamefont{Napolitano}},
  \bibinfo{author}{\bibfnamefont{J.}~\bibnamefont{Weiner}},
  \bibinfo{author}{\bibfnamefont{C.~J.} \bibnamefont{Williams}},
  \bibnamefont{and} \bibinfo{author}{\bibfnamefont{P.~S.}
  \bibnamefont{Julienne}}, \bibinfo{journal}{Phys. Rev. Lett.}
  \textbf{\bibinfo{volume}{73}}, \bibinfo{pages}{1352} (\bibinfo{year}{1994}).

\bibitem[{\citenamefont{Pillet et~al.}(1997)\citenamefont{Pillet, Crubellier,
  A~Bleton, Dulieu, Nosbaum, Mourachko, and Masnou-Seeuws}}]{Pillet97}
\bibinfo{author}{\bibfnamefont{P.}~\bibnamefont{Pillet}}, \bibnamefont{et~al.},
  \bibinfo{journal}{J. Phys. B} \textbf{\bibinfo{volume}{30}},
  \bibinfo{pages}{2801} (\bibinfo{year}{1997}).

\bibitem[{\citenamefont{C\^{o}t\'{e} and Dalgarno}(1998)}]{Cote98}
\bibinfo{author}{\bibfnamefont{R.}~\bibnamefont{C\^{o}t\'{e}}}
  \bibnamefont{and} \bibinfo{author}{\bibfnamefont{A.}~\bibnamefont{Dalgarno}},
  \bibinfo{journal}{Phys. Rev. A} \textbf{\bibinfo{volume}{58}},
  \bibinfo{pages}{498} (\bibinfo{year}{1998}).

\bibitem[{\citenamefont{Bohn and Julienne}(1999)}]{Bohn99}
\bibinfo{author}{\bibfnamefont{J.~L.} \bibnamefont{Bohn}} \bibnamefont{and}
  \bibinfo{author}{\bibfnamefont{P.~S.} \bibnamefont{Julienne}},
  \bibinfo{journal}{Phys. Rev. A} \textbf{\bibinfo{volume}{60}},
  \bibinfo{pages}{414} (\bibinfo{year}{1999}).

\bibitem[{\citenamefont{Ko\v{s}trun et~al.}(2000)\citenamefont{Ko\v{s}trun,
  Mackie, C\^{o}t\'{e}, and Javanainen}}]{Kostrun00}
\bibinfo{author}{\bibfnamefont{M.}~\bibnamefont{Ko\v{s}trun}},
  \bibinfo{author}{\bibfnamefont{M.}~\bibnamefont{Mackie}},
  \bibinfo{author}{\bibfnamefont{R.}~\bibnamefont{C\^{o}t\'{e}}},
  \bibnamefont{and}
  \bibinfo{author}{\bibfnamefont{J.}~\bibnamefont{Javanainen}},
  \bibinfo{journal}{Phys. Rev. A} \textbf{\bibinfo{volume}{62}},
  \bibinfo{pages}{063616} (\bibinfo{year}{2000}).

\bibitem[{\citenamefont{Timmermans et~al.}(1999)\citenamefont{Timmermans,
  Tommasini, Hussein, and Kerman}}]{Timmermans99}
\bibinfo{author}{\bibfnamefont{E.}~\bibnamefont{Timmermans}},
  \bibinfo{author}{\bibfnamefont{P.}~\bibnamefont{Tommasini}},
  \bibinfo{author}{\bibfnamefont{M.}~\bibnamefont{Hussein}}, \bibnamefont{and}
  \bibinfo{author}{\bibfnamefont{A.}~\bibnamefont{Kerman}},
  \bibinfo{journal}{Phys. Rep.} \textbf{\bibinfo{volume}{315}},
  \bibinfo{pages}{199} (\bibinfo{year}{1999}).

\bibitem[{\citenamefont{Goral et~al.}(2001)\citenamefont{Goral, Gajda, and
  Rzazewski}}]{Goral01}
\bibinfo{author}{\bibfnamefont{K.}~\bibnamefont{Goral}},
  \bibinfo{author}{\bibfnamefont{M.}~\bibnamefont{Gajda}}, \bibnamefont{and}
  \bibinfo{author}{\bibfnamefont{K.}~\bibnamefont{Rzazewski}},
  \bibinfo{journal}{Phys. Rev. Lett.} \textbf{\bibinfo{volume}{86}},
  \bibinfo{pages}{1397} (\bibinfo{year}{2001}).

\bibitem[{\citenamefont{Holland et~al.}(2001)\citenamefont{Holland, Park, and
  Walser}}]{Holland01}
\bibinfo{author}{\bibfnamefont{M.}~\bibnamefont{Holland}},
  \bibinfo{author}{\bibfnamefont{J.}~\bibnamefont{Park}}, \bibnamefont{and}
  \bibinfo{author}{\bibfnamefont{R.}~\bibnamefont{Walser}},
  \bibinfo{journal}{Phys. Rev. Lett.} \textbf{\bibinfo{volume}{86}},
  \bibinfo{pages}{1915} (\bibinfo{year}{2001}).

\bibitem[{\citenamefont{Javanainen and Mackie}(2002)}]{Javanainen02}
\bibinfo{author}{\bibfnamefont{J.}~\bibnamefont{Javanainen}} \bibnamefont{and}
  \bibinfo{author}{\bibfnamefont{M.}~\bibnamefont{Mackie}},
  \bibinfo{journal}{Phys. Rev. Lett.} \textbf{\bibinfo{volume}{88}},
  \bibinfo{pages}{090403} (\bibinfo{year}{2002}).

\bibitem[{\citenamefont{Fedichev et~al.}(1996)\citenamefont{Fedichev, Kagan,
  Shlyapnikov, and Walraven}}]{Fedichev96}
\bibinfo{author}{\bibfnamefont{P.~O.} \bibnamefont{Fedichev}},
  \bibinfo{author}{\bibfnamefont{Y.}~\bibnamefont{Kagan}},
  \bibinfo{author}{\bibfnamefont{G.~V.} \bibnamefont{Shlyapnikov}},
  \bibnamefont{and} \bibinfo{author}{\bibfnamefont{J.~T.~M.}
  \bibnamefont{Walraven}}, \bibinfo{journal}{Phys. Rev. Lett.}
  \textbf{\bibinfo{volume}{77}}, \bibinfo{pages}{2913} (\bibinfo{year}{1996}).

\bibitem[{\citenamefont{Bohn and Julienne}(1997)}]{Bohn97}
\bibinfo{author}{\bibfnamefont{J.~L.} \bibnamefont{Bohn}} \bibnamefont{and}
  \bibinfo{author}{\bibfnamefont{P.~S.} \bibnamefont{Julienne}},
  \bibinfo{journal}{Phys. Rev. A} \textbf{\bibinfo{volume}{56}},
  \bibinfo{pages}{1486} (\bibinfo{year}{1997}).

\bibitem[{\citenamefont{Javanainen and Mackie}(1998)}]{Javanainen98}
\bibinfo{author}{\bibfnamefont{J.}~\bibnamefont{Javanainen}} \bibnamefont{and}
  \bibinfo{author}{\bibfnamefont{M.}~\bibnamefont{Mackie}},
  \bibinfo{journal}{Phys. Rev. A} \textbf{\bibinfo{volume}{58}},
  \bibinfo{pages}{R789} (\bibinfo{year}{1998}).

\bibitem[{\citenamefont{Gerton et~al.}(2001)\citenamefont{Gerton, Frew, and
  Hulet}}]{Gerton01}
\bibinfo{author}{\bibfnamefont{J.~M.} \bibnamefont{Gerton}},
  \bibinfo{author}{\bibfnamefont{B.~J.} \bibnamefont{Frew}}, \bibnamefont{and}
  \bibinfo{author}{\bibfnamefont{R.~G.} \bibnamefont{Hulet}},
  \bibinfo{journal}{Phys. Rev. A} \textbf{\bibinfo{volume}{64}},
  \bibinfo{pages}{053410} (\bibinfo{year}{2001}).

\bibitem[{\citenamefont{McKenzie et~al.}(2002)\citenamefont{McKenzie,
  Hecker~Denschlag, H\"{a}ffner, Browaeys, Araujo, Fatemi, Jones, Simsarian,
  Cho, Simoni et~al.}}]{McKenzie02}
\bibinfo{author}{\bibfnamefont{C.}~\bibnamefont{McKenzie}},
  \bibnamefont{et~al.}, \bibinfo{journal}{Phys. Rev. Lett.}
  \textbf{\bibinfo{volume}{88}}, \bibinfo{pages}{120403}
  (\bibinfo{year}{2002}).

\bibitem[{\citenamefont{Drag et~al.}(2000)\citenamefont{Drag, Tolra, Dulieu,
  Comparat, Vatasescu, Boussen, Guibal, Crubellier, and Pillet}}]{Drag00}
\bibinfo{author}{\bibfnamefont{C.}~\bibnamefont{Drag}}, \bibnamefont{et~al.},
  \bibinfo{journal}{IEEE J. Quantum Electron.} \textbf{\bibinfo{volume}{36}},
  \bibinfo{pages}{1378} (\bibinfo{year}{2000}).

\bibitem[{\citenamefont{Schl\"{o}der et~al.}(2002)\citenamefont{Schl\"{o}der,
  Silber, Deuschle, and Zimmermann}}]{Schloder02}
\bibinfo{author}{\bibfnamefont{U.}~\bibnamefont{Schl\"{o}der}},
  \bibinfo{author}{\bibfnamefont{C.}~\bibnamefont{Silber}},
  \bibinfo{author}{\bibfnamefont{T.}~\bibnamefont{Deuschle}}, \bibnamefont{and}
  \bibinfo{author}{\bibfnamefont{C.}~\bibnamefont{Zimmermann}},
  \bibinfo{journal}{Phys. Rev. Lett} \textbf{\bibinfo{volume}{66}},
  \bibinfo{pages}{061403} (\bibinfo{year}{2002}).

\bibitem[{\citenamefont{Sackett et~al.}(1997)\citenamefont{Sackett, Bradley,
  Welling, and Hulet}}]{Sackett97b}
\bibinfo{author}{\bibfnamefont{C.~A.} \bibnamefont{Sackett}},
  \bibinfo{author}{\bibfnamefont{C.~C.} \bibnamefont{Bradley}},
  \bibinfo{author}{\bibfnamefont{M.}~\bibnamefont{Welling}}, \bibnamefont{and}
  \bibinfo{author}{\bibfnamefont{R.~G.} \bibnamefont{Hulet}},
  \bibinfo{journal}{Appl. Phys. B} \textbf{\bibinfo{volume}{65}},
  \bibinfo{pages}{433} (\bibinfo{year}{1997}).

\bibitem[{\citenamefont{Bradley et~al.}(1997)\citenamefont{Bradley, Sackett,
  and Hulet}}]{Bradley97a}
\bibinfo{author}{\bibfnamefont{C.~C.} \bibnamefont{Bradley}},
  \bibinfo{author}{\bibfnamefont{C.~A.} \bibnamefont{Sackett}},
  \bibnamefont{and} \bibinfo{author}{\bibfnamefont{R.~G.} \bibnamefont{Hulet}},
  \bibinfo{journal}{Phys. Rev. Lett.} \textbf{\bibinfo{volume}{78}},
  \bibinfo{pages}{985} (\bibinfo{year}{1997}).

\bibitem[{\citenamefont{Sackett et~al.}(1999)\citenamefont{Sackett, Gerton,
  Welling, and Hulet}}]{Sackett99}
\bibinfo{author}{\bibfnamefont{C.~A.} \bibnamefont{Sackett}},
  \bibinfo{author}{\bibfnamefont{J.~M.} \bibnamefont{Gerton}},
  \bibinfo{author}{\bibfnamefont{M.}~\bibnamefont{Welling}}, \bibnamefont{and}
  \bibinfo{author}{\bibfnamefont{R.~G.} \bibnamefont{Hulet}},
  \bibinfo{journal}{Phys. Rev. Lett.} \textbf{\bibinfo{volume}{82}},
  \bibinfo{pages}{876} (\bibinfo{year}{1999}).

\bibitem[{\citenamefont{Abraham
  et~al.}(1995{\natexlab{a}})\citenamefont{Abraham, Ritchie, McAlexander, and
  Hulet}}]{Abraham95b}
\bibinfo{author}{\bibfnamefont{E.~R.~I.} \bibnamefont{Abraham}},
  \bibinfo{author}{\bibfnamefont{N.~W.~M.} \bibnamefont{Ritchie}},
  \bibinfo{author}{\bibfnamefont{W.~I.} \bibnamefont{McAlexander}},
  \bibnamefont{and} \bibinfo{author}{\bibfnamefont{R.~G.} \bibnamefont{Hulet}},
  \bibinfo{journal}{J. Chem. Phys.} \textbf{\bibinfo{volume}{103}},
  \bibinfo{pages}{7773} (\bibinfo{year}{1995}{\natexlab{a}}).

\bibitem[{foo()}]{footnote}
\bibinfo{note}{In Ref. \cite{Gerton01}, we had erroneously assumed that $d_m$
  was quantized along the magnetic field direction, and considering laser
  polarization, overestimated $d_m$ by a factor of $\sqrt{2}$}.

\bibitem[{\citenamefont{Abraham
  et~al.}(1995{\natexlab{b}})\citenamefont{Abraham, McAlexander, Sackett, and
  Hulet}}]{Abraham95a}
\bibinfo{author}{\bibfnamefont{E.~R.~I.} \bibnamefont{Abraham}},
  \bibinfo{author}{\bibfnamefont{W.~I.} \bibnamefont{McAlexander}},
  \bibinfo{author}{\bibfnamefont{C.~A.} \bibnamefont{Sackett}},
  \bibnamefont{and} \bibinfo{author}{\bibfnamefont{R.~G.} \bibnamefont{Hulet}},
  \bibinfo{journal}{Phys. Rev. Lett.} \textbf{\bibinfo{volume}{74}},
  \bibinfo{pages}{1315} (\bibinfo{year}{1995}{\natexlab{b}}).

\bibitem[{\citenamefont{McAlexander et~al.}(1995)\citenamefont{McAlexander,
  Abraham, Ritchie, Williams, Stoof, and Hulet}}]{McAlexander95}
\bibinfo{author}{\bibfnamefont{W.~I.} \bibnamefont{McAlexander}},
  \bibinfo{author}{\bibfnamefont{E.~R.~I.} \bibnamefont{Abraham}},
  \bibinfo{author}{\bibfnamefont{N.~W.~M.} \bibnamefont{Ritchie}},
  \bibinfo{author}{\bibfnamefont{C.~J.} \bibnamefont{Williams}},
  \bibinfo{author}{\bibfnamefont{H.~T.~C.} \bibnamefont{Stoof}},
  \bibnamefont{and} \bibinfo{author}{\bibfnamefont{R.~G.} \bibnamefont{Hulet}},
  \bibinfo{journal}{Phys. Rev. A} \textbf{\bibinfo{volume}{51}},
  \bibinfo{pages}{R871} (\bibinfo{year}{1995}).

\end{thebibliography}

\newpage

\begin{figure}
\caption[Fig. 1]{Photoassociation resonance curve for $v' = 83$.
The solid circles are experimental data points and the solid line
is a Lorentzian fit to them.  The signal is the normalized number
of atoms remaining in the trap after the photoassociation pulse of
intensity 0.185 W/cm$^{2}$ and duration 2.0 ms. The initial number
of atoms is $5 \times 10^{5}$ at a temperature of $\sim$520 nK.
The uncertainty in relative frequency is $\sim$2 MHz, caused by
drift of the PA and reference laser frequencies. The data fit to a
rate constant with a Lorentzian width of 14 MHz, in good agreement
with the expected natural width of 11.7 MHz, which is twice the
atomic natural width \cite{Cote98}. The thermal contribution to
the lineshape is negligible at this temperature.}
\end{figure}

\begin{figure}[!]
\caption[Fig. 2]{Light-induced frequency shift of the
photoassociation resonance.  The solid circles are data.  The
error bars are uncertainty estimates for the identification of the
resonance peak positions.  The dashed line is a fit to a straight
line, where each data point is weighted by the inverse of the
error bar length.}
\end{figure}

\begin{figure}[!]
\caption[Fig. 3]{On-resonance photoassociation rate coefficient
$K_{p}$ as a function of laser intensity.  The open circles are
experimental data, obtained from trap loss analysis. The solid
line is the theoretical prediction, with no adjustable parameters.
The dashed line is an extrapolation of the theoretical result in
the low-intensity limit.}
\end{figure}

\newpage
\begin{figure}
\includegraphics[width=1\linewidth]{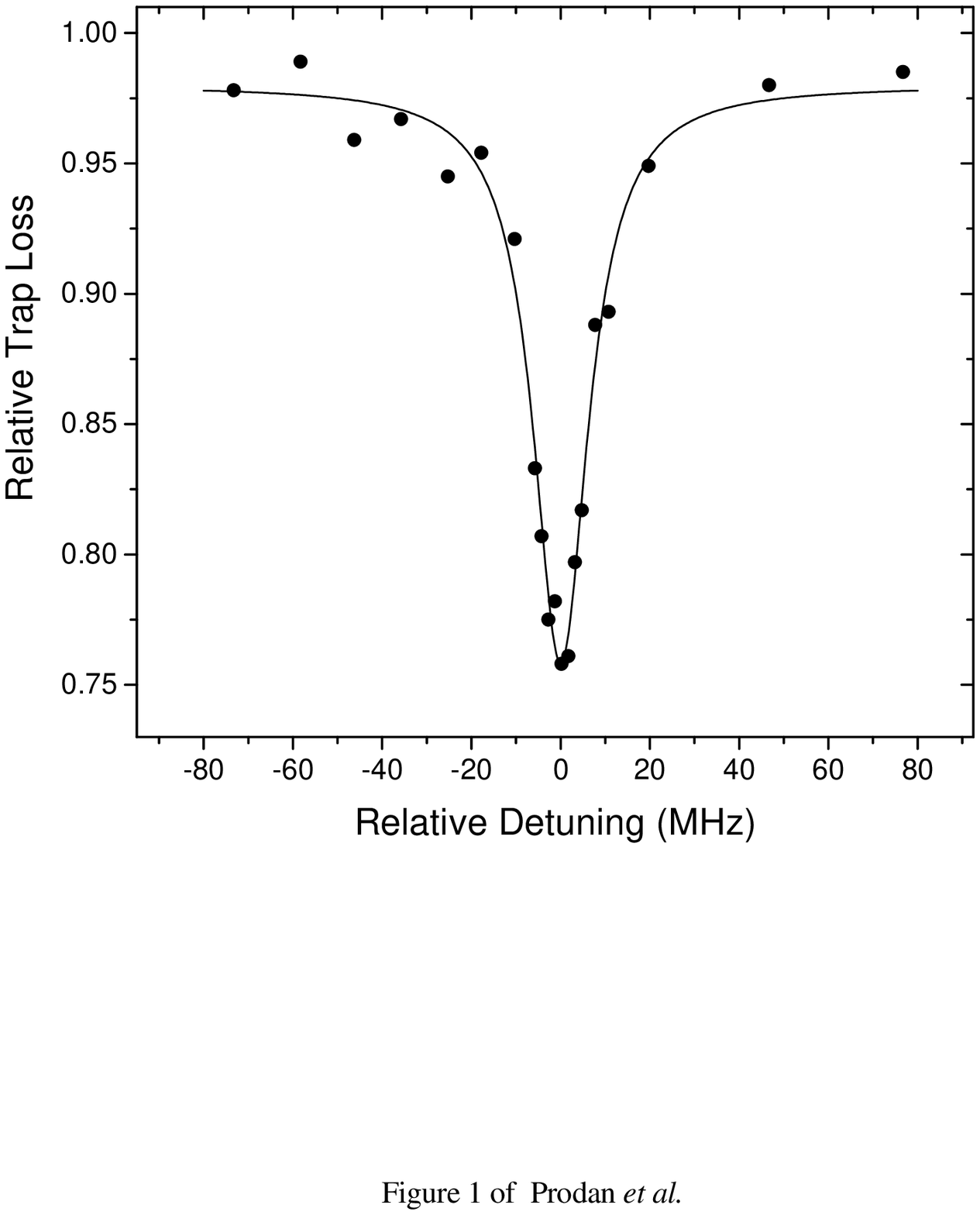}
\end{figure}

\newpage
\begin{figure}
\includegraphics[width=1\linewidth]{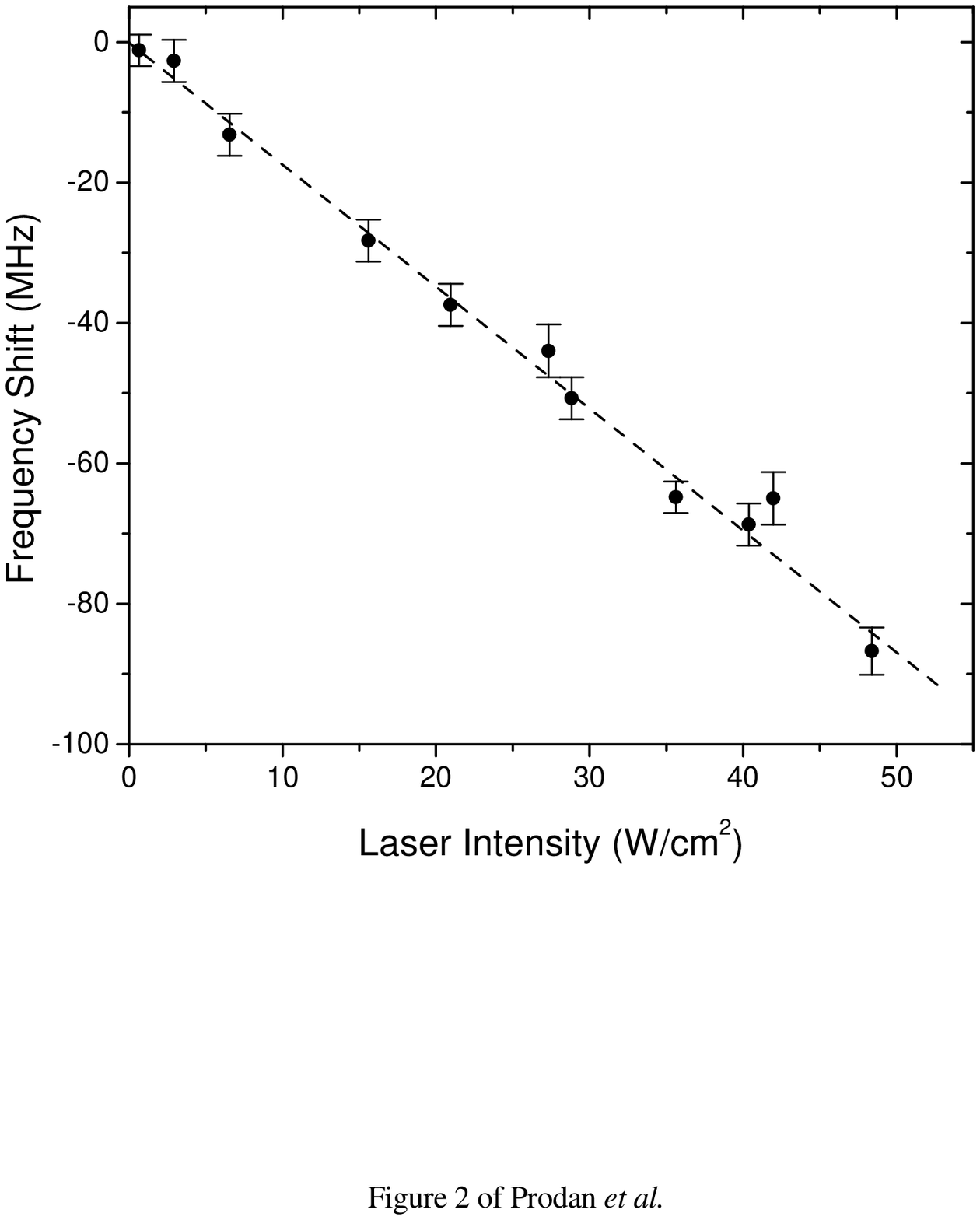}
\end{figure}

\newpage
\begin{figure}
\includegraphics[width=1\linewidth]{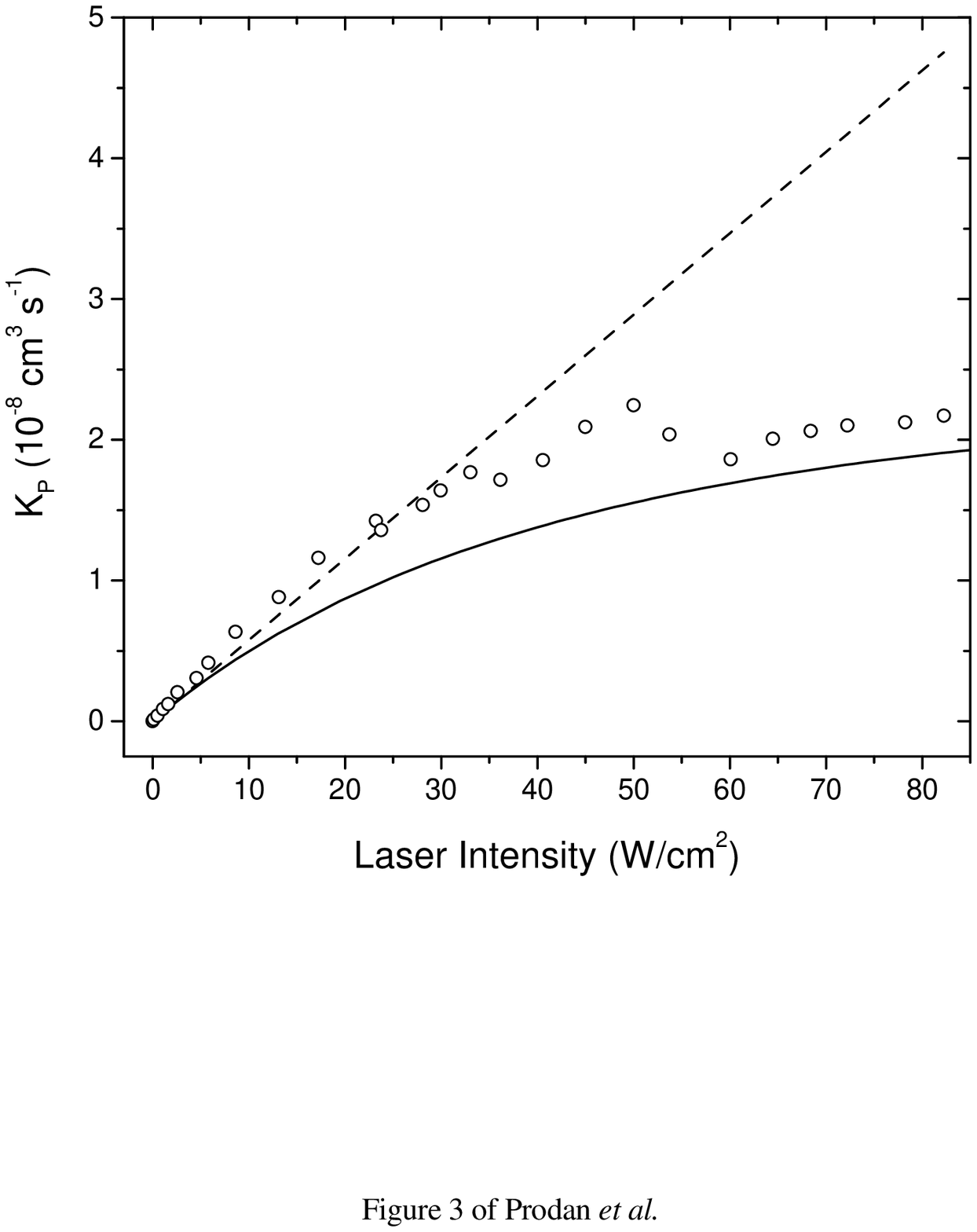}
\end{figure}

\end{document}